\documentclass[]{aa}
\usepackage{epsfig}
\usepackage{graphics}
\usepackage{txfonts}
\usepackage{subfigure}
\usepackage{natbib}
\usepackage{amsmath}
\usepackage{enumerate}
\usepackage{threeparttable}
\usepackage{caption}
\usepackage{ulem}
\usepackage[usenames,dvipsnames,svgnames,table]{xcolor}

\begin{document}

\title{3D hydrodynamical simulations of the impact of mechanical feedback on accretion in supersonic stellar-mass black holes}

\author{V. Bosch-Ramon\inst{1}}

\institute{Departament de F\'{i}sica Qu\`antica i Astrof\'{i}sica, Institut de Ci\`encies del Cosmos (ICC), Universitat de Barcelona (IEEC-UB), Mart\'{i} i Franqu\`es 1, E08028 Barcelona, Spain. \\\\  \email{vbosch@fqa.ub.edu}}


\titlerunning{3D simulations of mechanical feedback in supersonic stellar-mass black holes}

\abstract{Isolated stellar-mass black holes accrete gas from their surroundings, often at supersonic speeds, and can form outflows that may influence the accreted gas. The latter process, known as mechanical feedback, can significantly affect the accretion rate.} 
{We use hydrodynamical simulations to assess the impact of mechanical feedback on the accretion rate when the black hole moves supersonically through a uniform medium.} 
{We carried out three-dimensional (3D) hydrodynamical simulations of outflows fueled by accretion that interact with a uniform medium, probing scales equivalent to and larger than the accretor gravitational sphere of influence. In the simulations, the accretor is at rest and the medium moves at supersonic speeds. The outflow power is assumed to be proportional to the accretion rate. The simulations were run for different outflow-medium motion angles and velocity ratios. We also investigated the impact of different degrees of outflow collimation, accretor size, and resolution.} 
{In general, the accretion rate is significantly affected by mechanical feedback. There is a minor reduction in accretion for outflows perpendicular to the medium motion, but the reduction quickly becomes more significant for smaller angles. Moreover, the decrease in accretion becomes greater for smaller medium-to-outflow velocity ratios. On the other hand, the impact of outflow collimation seems moderate. Mechanical feedback is enhanced when the accretor size is reduced. For a population of black holes with random outflow orientations, the average accretion rate drops by (high--low resolution) $\sim 0.2-0.4$ and $\sim 0.1-0.2$ for medium-to-outflow velocity ratios of $1/20$ and $1/100$, respectively, when compared to the corresponding cases without outflow.} 
{Our results strongly indicate that on the considered scales, mechanical feedback can easily reduce the energy available from supersonic accretion by at least a factor of a few. This aspect should be taken into account when studying the mechanical, thermal, and non-thermal output of isolated black holes.}

\maketitle

\section{Introduction}\label{intro}

Wandering stellar-mass black holes that lack non-degenerate stellar companions (referred to as isolated black holes, IBH, hereafter, regardless of their multiplicity) are naturally expected to be present in the Galaxy simply based on studies of stellar evolution.  The
predicted numbers of these objects in the Galactic disk are as high as $\sim 10^8$ \citep[e.g.,][]{sha83} and, thus, significant
numbers of IBH ought to be present in all kinds of galactic environments. In fact, the presence of IBH in the Universe has been
indirectly inferred from gravitational lensing studies \citep[e.g.,][]{wyr16,kar20,wyr20} and directly found through the
detection of gravitational waves from black-hole binary mergers by the LIGO and Virgo collaborations
\citep[e.g.,][]{lig16,lig20}. Moreover, the presence of IBH have also been inferred through their effects on the
dynamical evolution of the galactic structures in which they are embedded \citep[see, e.g.,][and references therein]{gie21}.
However, despite theoretical expectations of their numbers and emission
\citep[e.g.,][]{mes75,mcd85,cam93,fuj98,ago02,mac05,bar12b,fen13,iok17,tsu18,mat18,tsu19,kim21}, thus far there has not been any electromagnetic radiation detected unambiguously from these
objects. It is worth noting that in addition to IBHs of stellar origin, black holes of stellar masses formed in the
very early universe, known as primordial black holes, may also exist \citep[e.g.,][]{zel67,haw71,car74,cha75}. If this were the case, these objects would make up
some fraction of the dark matter and could be responsible for (at least some) gravitational wave detections
\citep[e.g.,][]{bir16,cle16,sas16,sas18,del21}, leaving an imprint on the cosmic microwave background and the 21-cm Hydrogen
signals \citep[e.g.,][]{ric08,ali17,pou17,ber17,nak18,ber18,hek18,men19,hut19} and potentially populating the Galaxy in
significant numbers while emitting in different wavelengths \citep[e.g.,][]{gag17,man19,bos20}. 

Regardless of their origin, IBHs accreting from the environment can launch outflows (winds, jets, or both) under very different conditions.
The presence of such outflows, which may be quite common, can modulate the accretion of gas by transferring momentum and energy to the
medium  on the scales of the IBH gravitational sphere of influence, or even farther. This process is known as mechanical feedback
and it does not require much outflow power to affect the gas far from the IBH, as even relatively weak outflows could interfere with the
supersonic accretion process. In fact, few numerical studies of illustrative cases \citep{li20,bos20} have been done, which suggest that
this interference may be rather strong. Therefore, the potential ubiquity of outflows, and a possible high efficiency of mechanical
feedback, may severely reduce the energetics of IBH accretion already relatively far from the IBH. In this way, mechanical feedback would
affect, for instance, the radiation produced by accretion or the outflows themselves, the potential non-thermal particle production, IBH
growth, or effects on matter and radiation in the present and the early Universe \cite[see,
e.g.,][]{sok16,iok17,lev18,zei19,gru19,li20,bos20,tak21,tak21b}. 

It is still not clear at a quantitative level how the accretion rate depends on the outflow orientation and collimation degree, or the
accretor-medium relative velocity, as a broad exploration of mechanical feedback in accreting supersonic IBH has not been undertaken thus far.
Attempting to improve our understanding of the role of mechanical feedback in accreting supersonic IBHs, we performed, for the first
time, a wide exploration of mechanical feedback in stellar-mass IBH through a set of numerical three-dimensional (3D) simulations, probing
different outflow orientations and medium velocities, outflow opening angles, and the effect of simulation limitations such as
resolution or the accretor size. We focused on the effect of mechanical feedback on the accretion rate on scales similar to or larger
than the IBH sphere of influence. 

The article is organized as follows. In Sect.~\ref{phy}, we describe the studied scenario and
present a simplified analysis of mechanical feedback that helped to guide the numerical
calculations. Then, the numerical simulations are introduced in Sect.~\ref{sim}, and their
results are presented and discussed in Sects.~\ref{res} and \ref{dis}, respectively. Throughout
this work, the convention $Q_b = Q / 10^b$ is adopted, with $Q$ given in cgs units (unless
stated otherwise).

\section{Physical scenario}\label{phy}

\subsection{Accretion and outflow}
 
The scenario investigated in this work consists of an IBH surrounded by a uniform medium with velocity $v_{\rm IBH}$, which is
taken well above the sound speed of the medium. The IBH produces a bipolar, axisymmetric, and accretion-powered outflow
whose injection was modeled phenomenologically and located close to the IBH. We adopted a set of parameters for
which the interacting flows evolve adiabatically. 

The impact of the IBH outflow on the accretion rate was studied on scales similar to or larger than the sphere of influence of the IBH. This region defined around the IBH has a radius \citep{hoy39,bon44}:
\begin{equation} 
r_{\rm acc}\sim 2\, G M_{\rm IBH} v_{\rm IBH}^{-2} \approx 8\times 10^{13} M_{\rm
IBH,1.5}\, v_{\rm IBH,7}^{-2} \ {\rm cm}\,,
\label{eq:accretion_radius} 
\end{equation} 
and is much larger than the region where the outflow forms. As the outflow formation occurs much closer to the IBH, the escape velocity is expected to be $\gg v_{\rm IBH}$. Thus, the outflow velocity ($v_{\rm o}$) is also expected to be $\gg v_{\rm IBH}$.

In the absence of mechanical feedback effects, the accretion rate of a (highly) supersonic accretor can be characterized using $r_{\rm acc}$ via \citep{hoy39,bon44}:
\begin{equation}
\begin{aligned}
\dot{M}_{\rm IBH}&\approx \pi r_{\rm acc}^2 \rho_{\rm m} v_{\rm IBH} \approx
3.2\times 10^{11} M_{\rm IBH,1.5}^2\, n_{\rm m,0}\, v_{\rm IBH,6.5}^{-3} \ {\rm g\, s^{-1}}\\&\approx 
5.2\times 10^{-15}M_{\rm IBH,1.5}^2\, n_{\rm m,0}\, v_{\rm IBH,6.5}^{-3} \ {\rm M_\odot\, yr^{-1}},
\end{aligned}
\label{eq:bondi_hoyle_accretion_rate}
\end{equation}
where $n_{\rm m}$ is the medium density \citep[see, e.g., Eq. 32 in][for the case with an appreciable sound speed]{edg04}. It is worth mentioning, as discussed in \cite{bos20}, that the accretion rates expected in typical scenarios are far lower than the Eddington rate, which implies that accretion takes place through a radiatively inefficient, geometrically thick in-flowing structure.

The power of each side of the bipolar outflow is linked to the accretion rate via $L_{\rm o}=(1/4)\epsilon\dot{\bar
M}_{\rm IBH}v_{\rm o}^2$, where $\dot{\bar M}_{\rm IBH}$ is the accretion rate under the effects of mechanical feedback
and $\epsilon$ is a free parameter that determines the fraction of $\dot{\bar M}_{\rm IBH}$ that goes to the bipolar
outflow. We note that the power of the outflow, as it is defined, and the fact that the latter is initially highly
supersonic allow for the derivation of the outflow mass rate: $\dot{M}_{\rm o}=\epsilon\dot{\bar M}_{\rm IBH}$. As
demonstrated in Eqs. 10 and 11 in \cite{bos20}, the energetic requirement for an outflow to escape the ram pressure
of the in-falling material is quite modest, and we chose in this work a rather low value of 0.1 for $\epsilon$ . On the
other hand, \cite{bar12a} showed that the efficiency in outflow production under somewhat similar conditions (i.e., low
angular-momentum accretion) could be in fact as high as $\epsilon v_{\rm o}^2\lesssim c^2$, whereas what we take here  is $\epsilon v_{\rm o}^2\ll
c^2$. We chose a much lower (and arguably more conservative) value for this efficiency for two reasons: 1) to show that even relatively weak winds can significantly affect the accretion rate; 2) perhaps
even more importantly,  a modest $L_{\rm o}$-value allows for the use of a computational grid that is significantly
smaller than that needed in the case of a powerful outflow.

The outflow assumed here can still easily reach beyond $r_{\rm acc}$, allowing it to directly interact hydrodynamically
with material that is still not significantly affected by the IBH gravity. Through this mechanical feedback, the outflow
deposits momentum and energy on the incoming medium (as seen by the IBH), thereby compressing and heating the gas and
potentially reducing the accretion rate.

\subsection{Relevant dependencies of mechanical feedback}

\begin{figure}        
\includegraphics[width=8cm]{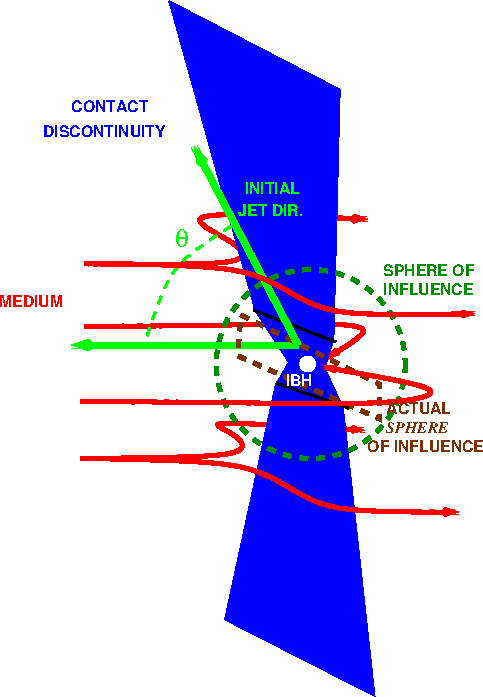}
\caption{Sketch of the scenario (not to scale). The relation between the green circumference and the smaller brown rectangle determines how much medium finally reaches the accretor. The thick red solid lines represent the medium streamlines that may or may not reach the accretor depending on their impact parameter.}
\label{sk}
\end{figure}

Mechanical feedback effectively reduces how much medium is captured (as illustrated in Fig.~\ref{sk}) and the actual accretion rate, $\dot{\bar M}_{\rm IBH}$, will be smaller than the value of $\dot{M}_{\rm IBH}$ without an outflow by a factor of: 
\begin{equation}
\xi=\dot{\bar M}_{\rm IBH}/\dot{M}_{\rm IBH}\sim z_{\rm acc}/r_{\rm acc}\,. 
\end{equation}
The value of $\xi$ should mostly depend on how much energy is injected by the outflow in the incoming medium, as this would modify the effective region of influence of the IBH. The injection of energy takes place at the contact discontinuity (CD) between the medium and the outflow, mediated by the transfer of some momentum from the outflow to the medium.

The rate at which the outflow transfers momentum to the medium at the CD can be roughly estimated from the outflow momentum rate component facing the medium, which in the non-relativistic case is:
\begin{equation}
\dot{p}_{\rm o\perp}\sim (1/2)\dot{M}_{\rm o}v_{\rm o}\cos\theta=(1/2)\epsilon\dot{\bar M}_{\rm IBH}v_{\rm o}\cos\theta\,,
\label{pperp}
\end{equation}
where $\theta$ is the angle that characterizes the outflow-medium motion angle (see Fig.~\ref{sk}). This equation assumes that the outflow and the medium interact obliquely and that the outflow becomes subsonic at the oblique shock imparted by the medium impact, in the shock frame. For cases with $\theta$ approaching $90^\circ$, further physical information would be required for a proper estimate of $\dot{p}_{\rm o\perp}$, as the opening angle of the outflow becomes important. On the other hand, for $\theta\sim 0$, the flows interact head-on and the situation would be very different. 

Assuming that all the medium within $r_{\rm acc}$ is intercepted by the CD and shocked, the amount of energy per time transferred to the medium shocked against the CD, within a certain distance from the accretor, can be roughly characterized as:
\begin{equation}
L_{\rm o\perp}\sim \dot{p}_{\rm o\perp}v_{\rm IBH}\,.
\end{equation}
This extra energy, pumped into the medium by getting obliquely impacted by the outflow, can lead to a very much enhanced medium sound speed. Assuming that the mass rate entering the region of interest is $\dot{\bar M}_{\rm IBH}$, we can estimate this sound speed via:
\begin{equation}
c_{\rm s}\sim\sqrt{\bar\gamma L_{\rm o\perp}/\dot{\bar M}_{\rm IBH}}=\sqrt{\bar\gamma\epsilon v_{\rm o}v_{\rm IBH}\cos\theta/2}\,.
\label{cs}
\end{equation} 
After being stopped by the CD, the shocked medium is initially subsonic, and thus $c_{\rm s}$ becomes a characteristic velocity of
the problem. 
For the aim of performing numerical simulations, a heuristic approach is adopted in what follows, in which $\xi$ is considered to be
mostly a function of $v_{\rm IBH}/c_{\rm s}$ and, thus, a function of $\eta_v$ and $\cos\theta$, where $\eta_v=v_{\rm IBH}/v_{\rm
o}$. 

As explained above, $\epsilon$ is fixed to 0.1, which is high enough to allow the outflow to reach beyond the IBH sphere of influence
\citep{bos20}. This is done under the assumption that significantly larger $\epsilon$-values would lead at least to a similar
impact on the part of mechanical feedback on accretion (if not more), but a devoted study is needed for a proper, quantitative assessment. Despite the
dependency of $\dot{M}_{\rm IBH}$ on $M_{\rm IBH}$ is rather strong, the previous analysis allows us to neglect this parameter as
the intent of this investigation is focused solely on relative comparisons. On the other hand, despite the outflow half-opening angle ($\chi$)
does not appear in Eq.~\ref{cs}, we studied two different values of $\chi$ in the simulations for an exploratory evaluation of
its role.

We focused on cases in which the interacting flows were adiabatic, and a fortunate consequence of this assumption is that $\xi$ is
independent of $n_{\rm m}$. We did not consider conditions under which the interacting flows would evolve in the radiative regime
due to the difficulties in generalizing the results in that case, as they are more difficult to scale and the  computing costs are higher. Additional complex phenomena are
expected to arise in the radiative regime, such as the formation of thin shocked gas shells, cooling-enhanced instabilities,
enhanced interaction-structure disruption, and so on \citep[see, e.g.,][]{zei19}. Although these phenomena would likely yield
quantitative differences in the effect of mechanical feedback with respect to the adiabatic case, their study has been set aside for a future
work. A related caveat is that without radiative cooling, accretion can be halted by pure adiabatic compression when the accreted gas becomes too hot close
enough to the IBH. In our case, given the relatively large accretor sizes considered, this effect is not significant. In a
way, this effect is included here by assuming the presence of an outflow, as the latter can be produced in the vicinity of an IBH
precisely due to an excess of thermal energy \citep[e.g.,][]{sad16}.

In the next sections, we explore numerically how accretion is affected by an outflow accounting for changes in $\theta$ and
$\eta_v$. Two different resolutions are also adopted, with the higher resolution being twice the lower one, allowing for the
study of cases with half the outflow half-opening angle and accretor size.

\section{Simulations}\label{sim}

The simulations done for this work were carried out in 3D, which allowed for the exploration of a much broader parameter space than
that studied in the 2D simulations of \cite{bos20}. Since the adopted $v_{\rm o}$ is $\ll c$, the performed numerical calculations
solved the non-relativistic equations of hydrodynamics in Cartesian coordinates for the conservation of mass, momentum, and energy,
using a code based on a finite-difference scheme that is of the third order in space and second order in time. For simplicity, the
simulated flows were assumed to be one-species, non-relativistic ideal gases that evolve adiabatically, with gas adiabatic index
$\hat\gamma=5/3$. We adopted the same Riemann solver and spatial discretization scheme as in \cite{bos20}: the Marquina flux
formula \citep{don96} and the PPM method, as presented in \cite{mig05}, respectively. For further details on the code employed, we refer to \cite{del17} and references therein. The employed code is parallelized through OpenMP, and the calculations
were done in a work station using ten processors. The computing time devoted to all the performed simulations was $\sim
2\times10^4$~CPU~hours.

The accretor was simulated as a quasi-spherical region in the computational grid with pressure and density much lower than in the
surroundings, within which the material is forced to move radially inwards at the local free-fall velocity. The accretor gravitational
force was introduced in the hydrodynamical equations as a source term \citep[see, e.g.,][]{tor09}. Through the poles of that spherical
region, a radial outflow symmetric around the $\hat z$-direction enters into the simulation region with half-opening angle, $\chi$,
a velocity of $v_{\rm o}=10^9$~cm~s$^{-1}$, and Mach number $\approx 8$. The Mach number is high enough to mimic the evolution of a flow
that is cold, as expected that far from the outflow launching region. Since the outflow is initially in overpressure with the
surroundings, its Mach number increases further outside the outflow inlet. The value of $\chi$ was set to 0.4 (radians; i.e., $\approx 23^\circ$) in all simulation
runs but in one case, in which a $\chi$-value of 0.2 was adopted to probe the effects of a more collimated outflow. Initially, in all
simulations, outside the accretor the medium initial density is $\rho_{\rm m}=n_{\rm m}m_{\rm H}=m_{\rm H}$~g~cm$^{-3}$, where $m_{\rm
H}\approx 1.7\times 10^{-24}$~g is the Hydrogen atom mass. Two different initial medium velocities were studied: $v_{\rm IBH}= 10^7$
and $5\times 10^7$~cm~s$^{-1}$, with a Mach number of 3. The grid boundary conditions were set to in-flow at the boundaries where the medium entered the computational domain, reflection at $y=0$, and free-flow in the remaining boundaries. 

The gravitational force of the IBH was modeled as generated by a point-like object with mass $M_{\rm IBH}=30$~M$_\odot$
located at $(0,0,0)$. The sphere of influence of such an object has a radius $r_{\rm acc}\approx 3\times 10^{12}$~cm for
$v_{\rm IBH}=5\times 10^7$~cm~s$^{-1}$, and $\approx 8\times 10^{13}$~cm for $v_{\rm IBH}=10^7$~cm~s$^{-1}$ (see
Eq.~\ref{eq:accretion_radius}). The corresponding accretion timescales are $t_{\rm acc}=r_{\rm acc}/v_{\rm IBH}\approx
6.4\times 10^4$~s and $\approx 8\times 10^6$~s for $v_{\rm IBH}=5\times 10^7$~cm~s$^{-1}$ and $v_{\rm
IBH}=10^7$~cm~s$^{-1}$, respectively.

\subsection{Studied cases}

For the simulations with $v_{\rm IBH}=5\times 10^7$~cm~s$^{-1}$ and low resolution (called reference cases), the considered values of $\theta$ were
$90^\circ$, $60^\circ$, $45^\circ$, $30^\circ$, and $0^\circ$, whereas for $v_{\rm IBH}=10^7$~cm~s$^{-1}$, they were
$90^\circ$, $45^\circ$, and $0^\circ$. For $v_{\rm IBH}=5\times 10^7$~cm~s$^{-1}$ and $\theta=60^\circ$ and $45^\circ$,
runs were carried out with twice the resolution of the reference cases, and additionally for $\theta=60^\circ$, two different accretor sizes and $\chi$-values were explored. All the listed parameter configurations are presented
in Table~\ref{tab}. 

In addition to the simulations of the described cases, several test trials were run to check the robustness of the
simulations, in particular for runs with the longest computing times. For instance, the simulations with $v_{\rm IBH}= 10^7$~cm~s$^{-1}$ were
approximately $\sim 20$ times longer than the corresponding reference cases, which made difficult to use large grids.
Therefore, the behavior of $\dot{\bar M}_{\rm IBH}$ was checked with relatively shorter simulations involving larger grids
and $\approx 2/3$ of the resolution. This was done for comparison for the cases with $\theta=45^\circ$ and $v_{\rm IBH}=
10^7$~cm~s$^{-1}$ as well as $v_{\rm IBH}=5\times 10^7$~cm~s$^{-1}$ and, for completeness, for the cases with $v_{\rm IBH}=
10^7$~cm~s$^{-1}$ and $\theta=30^\circ$ and $60^\circ$ (to probe a slightly different region of the parameter space).
These test results yielded quite similar results (not shown here) to those obtained in the main simulations, giving 
 us confidence on the soundness of the numerical solutions, particularly regarding the accretor, outflow inlet,
and grid sizes.

\begin{table}
 \begin{center}
        \caption{List of the (free) parameter values used in the simulations.}
        \begin{tabular}{l c c c c c}
    \hline 
Low res. ($l^{\rm uni}_{\rm cell}=r_{\rm acc}/30$) & & & & & \\    
    \hline
$\theta$: & $90^\circ$ & $60^\circ$ & $45^\circ$ & $30^\circ$ & $0^\circ$ \\      
$\chi$ [radians]: & 0.4 & 0.4 & 0.4 & 0.4 & 0.4 \\                  
$v_{\rm IBH}$ [$\times 10^7$~cm~s$^{-1}$]: & 1, 5 & 5  & 1, 5  & 5  & 1, 5 \\    
$\mathcal{R}$: & $1/3$ & $1/3$& $1/3$& $1/3$& $1/3$ \\
\hline
High res. ($l^{\rm uni}_{\rm cell}=r_{\rm acc}/60$) & & & & & \\    
\hline
$\theta$: & & $60^\circ$ & $45^\circ$ & & \\
$\chi$ [radians] & & 0.2,0.4 & 0.4 & & \\                  
$v_{\rm IBH}$ [$\times 10^7$~cm~s$^{-1}$]:  & & 5 &  5 & & \\   
$\mathcal{R}$: & & $1/6$,$1/3$ & $1/6$  &  \\    
   \hline
    \end{tabular}
    \label{tab}
 \end{center}
\end{table}

\begin{table}
 \begin{center}
        \caption{Number of cells for the whole grid and for the uniform grid region.}
        \begin{tabular}{l c c }
    \hline 
$l^{\rm uni}_{\rm cell}=r_{\rm acc}/30$ (all $\theta$) & $v_{\rm IBH}=5\times 10^7,$ & $10^7$~cm~s$^{-1}$ \\    
    \hline
$N_x,N^{\rm uni}_x$: & 200,20 & 185,20 \\
$N_y,N^{\rm uni}_x$: & 100,10 & 100,10 \\
$N_z,N^{\rm uni}_x$: & 200,20 & 270,20 \\
\hline
$l^{\rm uni}_{\rm cell}=r_{\rm acc}/60$ ($\theta=60^\circ,45^\circ$) &  &  \\    
\hline
$N_x,N^{\rm uni}_x$: & 260,40 & \\
$N_y,N^{\rm uni}_y$: & 130,20 & \\
$N_z,N^{\rm uni}_z$: & 260,40 & \\
    \hline
    \end{tabular}
    \label{tab2}
 \end{center}
\end{table}

\subsection{Computational grid}

The computational grid is inhomogeneous. In the low-resolution runs, the sphere of influence has a radius of $\sim
30$ cells, and the accretor region radius is 10 cells. In the high-resolution runs, the sphere of influence has a
radius of $\sim 60$ cells and the accretor radii are 10 or 20 cells, depending on the studied case. A ratio of accretor radius to $r_{\rm acc}$
($\mathcal{R}$; see Table.~\ref{tab}) of $1/3$ is not optimal for accuracy, but is enough for semi-quantitative
estimates and was generally adopted because it allowed us to run a larger number of simulations. Taking $\mathcal{R}=1/6$ yields more accurate results as long as the accretor is reasonably resolved ($\sim 10$~cells seemed to
work well enough for our purposes); \cite{bos20} found a reasonable level of convergence around
$\mathcal{R}\approx 1/6$, whereas \citealt{zei19} found an error of $\approx 10$\% for $\mathcal{R}=1/10$ in their accretion test
calculations. In the high- (low-) resolution runs, an accretor radius of 10 cells yields
$\mathcal{R}=1/6$ (1/3). The minimum $\chi$
was set such that the outflow diameter at injection was $\approx 8$~cells, which was the case in all the simulations except the one with high resolution and $\mathcal{R}=1/3$, for which it was $\approx 16$ cells. 
An accretor radius of 20 (10) cells and minimum outflow diameter of 8~cells allowed a minimum $\chi$ of 0.2 (0.4).
Systematic errors for an outflow inlet with radius $\approx 8$~cells are expected to be of $\sim 10$\% in the outflow mass, momentum, and energy rates, sufficient for the
intended level of approximation. 

For the reference cases, the whole computational grid consists of $200\times100\times200$ ($N_x\times N_y\times N_z$)
cells, ranging from $x,z\approx -8.3$ to $8.3\,r_{\rm acc}$, and from $y=0$ to $8.3\,r_{\rm acc}$; a run without outflow
was also done with $z>0$ and the upper half of the volume. The grid includes a uniform region centered at $(0,0,0)$ with
volume $20\times10\times20$ ($N^{\rm uni}_x\times N^{\rm uni}_y\times N^{\rm uni}_z$) cubic cells with size $l^{\rm
uni}_{\rm cell}=r_{\rm acc}/30$, fully encompassing the accretor and the outflow inlet. 

In the runs with $v_{\rm IBH}=10^7$~cm~s$^{-1}$, the grid ranges from $x\approx -2.5$ to $17.5\,r_{\rm acc}$, from $y=0$
to $8.75\,r_{\rm acc}$, and from $z\approx -17.5$ to $17.5\,r_{\rm acc}$. The uniform grid has the same $N^{\rm uni}_x$, $N^{\rm
uni}_y$, $N^{\rm uni}_z$ values as in the case with $v_{\rm IBH}=5\times 10^7$~cm~s$^{-1}$, whereas now $N_x=190$,
$N_y=100,$ and $N_z=270$. 

In the high-resolution simulations, $v_{\rm IBH}$ was again set to $5\times 10^7$~cm~s$^{-1}$, and two values for 
$\theta$, $60^\circ$ and $45^\circ$, were considered. The grid now ranges from $x,z\approx -6.75$ to $6.75\,r_{\rm
acc}$, and from $y=0$ to $6.75\,r_{\rm acc}$. The whole grid has $N_x=260$, $N_y=130$ and $N_z=260$, and the central
uniform region has $N^{\rm uni}_x=40$, $N^{\rm uni}_y=20,$ and $N^{\rm uni}_z=40$, and is made of cubic cells with $l^{\rm
uni}_{\rm cell}=r_{\rm acc}/60$. In the case with $\theta=60^\circ$, three different cases were studied probing
different accretor sizes and outflow opening angles: $\mathcal{R}=1/3$ ($\chi=0.4$) and $1/6$ ($\chi=0.4$) and
$\chi=0.2$ ($\mathcal{R}=1/3$). In the case with $\theta=45^\circ$, the parameter values were the same as in the
corresponding reference case, but with $\mathcal{R}$, now set to $1/6$. 

The non-uniform grid region is made of cells expanding at a fix rate of 1.019 per spatial step, starting beyond cell $N_x^{\rm uni}$,
$N^{\rm uni}_y$, and $N^{\rm uni}_z$ independently in each direction. To avoid numerical artifacts, the cell aspect ratio was kept
below $\sim 10$ in the whole grid \citep{per05}, and the maximum aspect ratio was only achieved at the most peripheric grid regions.
The information on the number of cells for the uniform and the whole grid in each simulation is given in Table~\ref{tab2}.

The value of $\dot{\bar M}_{\rm IBH}$ was computed two cells farther from the accretor boundary in all cases but in the high-resolution case with $\mathcal{R}=1/3$, for which it was four cells. We recall here that the power of each component of the
bipolar outflow is related to $\dot{\bar M}_{\rm IBH}$ via $L_{\rm o}=0.025\,\dot{\bar M}_{\rm IBH}v_{\rm o}^2$.
Given that $v_{\rm o}=10^9$~cm~s$^{-1}$, we may alternatively write $L_{\rm o}=2.8\times 10^{-5}\,\dot{\bar M}_{\rm
IBH}c^2$, which indicates that the adopted outflow power is rather small when compared to what is potentially available
close to the IBH event horizon, on the order of $\sim \dot{\bar M}_{\rm IBH}c^2$ \citep[see][and references therein]{bar12a}.

\section{Results}\label{res}

\begin{figure}        
       \centering       
       \includegraphics[width=8cm]{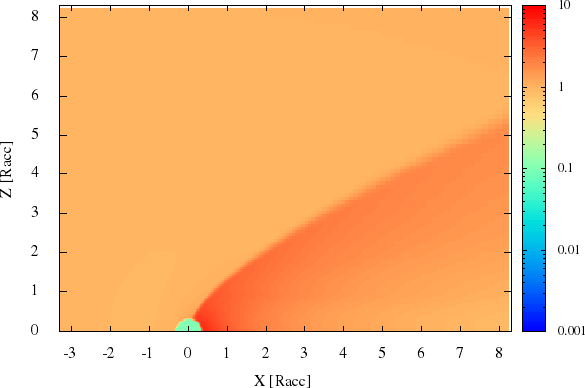}
       \includegraphics[width=8cm]{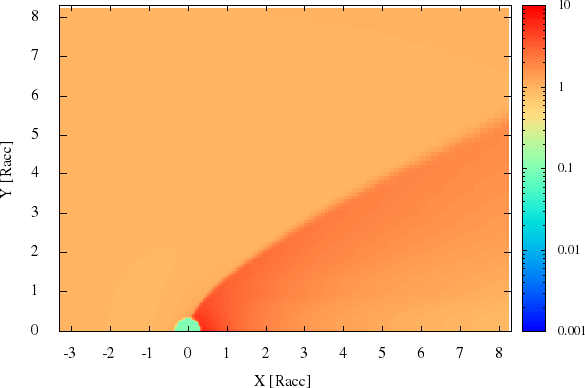}
       \caption{Density map on the $y=0$ (top panel) and $z=0$ planes (bottom panel) in the case without outflow, at $t_{\rm sim}\approx 23\,t_{\rm acc}$. The initial medium velocity and density are $5\times 10^7$~cm~s$^{-1}$ and $1.7\times 10^{-24}$~cm$^{-3}$, respectively, and $\mathcal{R}=1/3$ and $M_{\rm IBH}=30$~M$_\odot$. The axis units are $r_{\rm acc}\approx 3\times 10^{12}$~cm, and the color scale is normalized to the initial medium density.
}
       \label{densm1}
\end{figure}

\begin{figure}        
       \centering       
       \includegraphics[width=8cm]{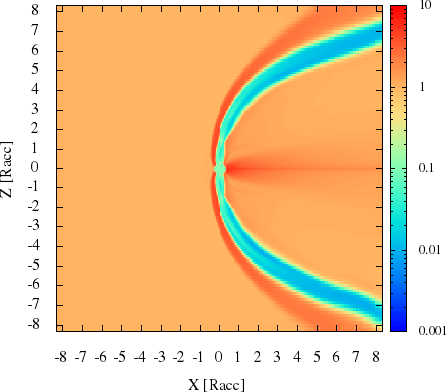}
       \includegraphics[width=8cm]{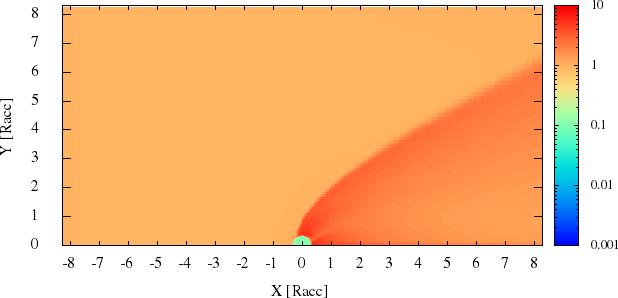}
       \caption{Density map on the $y=0$ (top panel) and $z=0$ planes (bottom panel) for an outflow with $\chi=0.4$ and $\theta=90^\circ$, at $t_{\rm sim}\approx 23\,t_{\rm acc}$. The initial medium velocity is $5\times 10^7$~cm~s$^{-1}$, and $\mathcal{R}=1/3$. The axis units are $r_{\rm acc}\approx 3\times 10^{12}$~cm.}
       \label{densm2}
\end{figure}

\begin{figure}        
       \centering      
       \includegraphics[width=8cm]{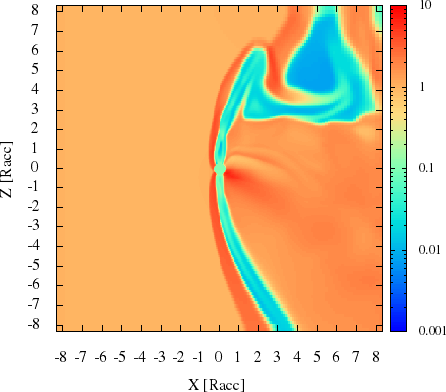}
       \includegraphics[width=8cm]{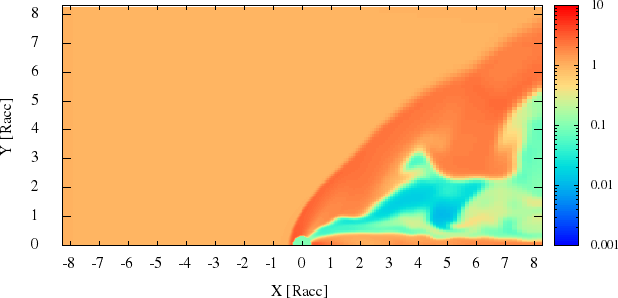}      
       \caption{Same as in Fig.~\ref{densm2} but for $\theta=60^\circ$.}
       \label{densm3}
\end{figure}

\begin{figure}        
       \centering       
       \includegraphics[width=8cm]{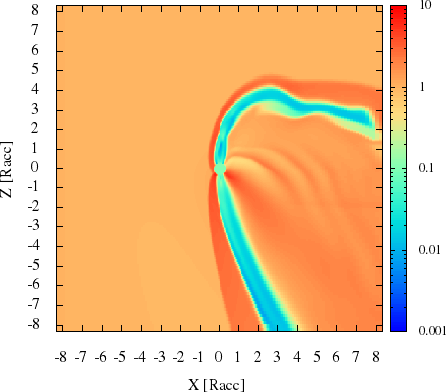}
       \includegraphics[width=8cm]{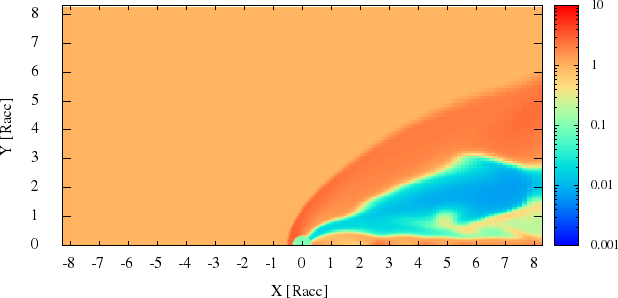}
       \caption{Same as in Fig.~\ref{densm2} but for $\theta=45^\circ$.}
       \label{densm4}
\end{figure}

\begin{figure}        
       \centering       
       \includegraphics[width=8cm]{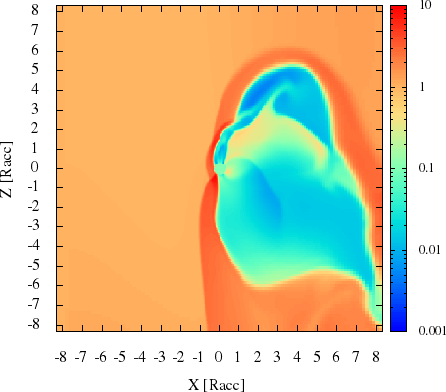}
       \includegraphics[width=8cm]{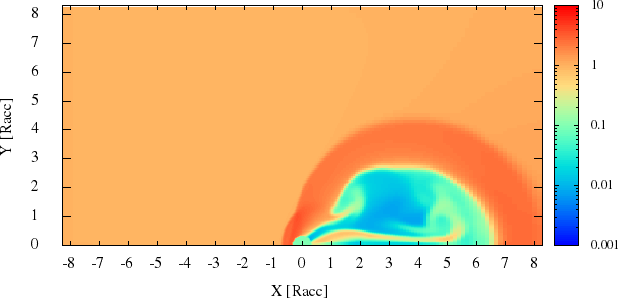}
       \caption{Same as in Fig.~\ref{densm2} but for $\theta=30^\circ$.}
       \label{densm5}
\end{figure}

\begin{figure}        
       \centering       
       \includegraphics[width=8cm]{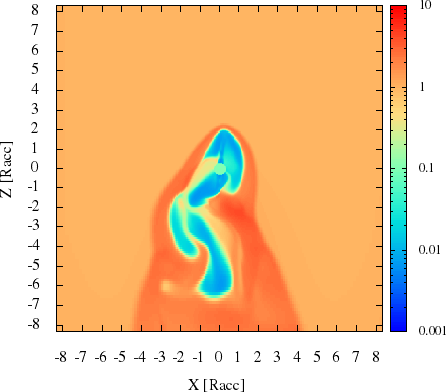}
       \includegraphics[width=8cm]{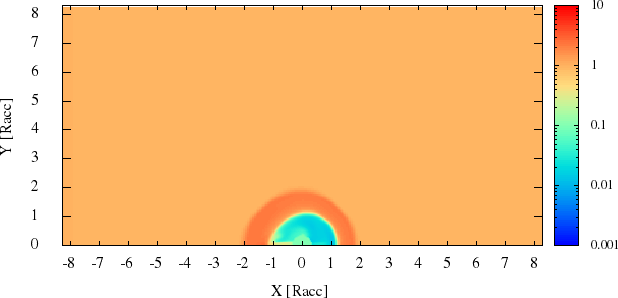}      
       \caption{Same as in Fig.~\ref{densm2} but for $\theta=0^\circ$.}
       \label{densm6}
\end{figure}

\begin{figure}        
\includegraphics[width=8cm]{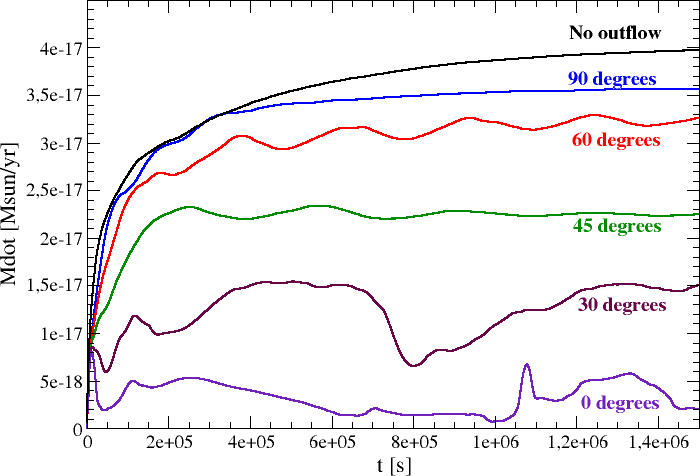}
\caption{Accretion rate for the reference cases (from top to bottom): without outflow and with outflow and $\theta=90^\circ$ (perpendicular medium and outflow), $60^\circ$, $45^\circ$, $30^\circ$, and $0^\circ$ (parallel medium and outflow). The results are for simulation times up to $t_{\rm sim}\approx 23\,t_{\rm acc}$, where $t_{\rm acc}\approx 6.4\times 10^4$~s.}
\label{accr}
\end{figure}

We show in Figs.~\ref{densm1}-\ref{densm6} the density maps for the reference cases at the end of the runs (time $t_{\rm sim}$), in the $y=0$ and $z=0$ planes and with no outflow case included. The computation time, $t_{\rm sim}$, was chosen such that the simulations were run
until they reached a quasi-steady (QS) state. This means that they ran long enough for the shocked external medium (the slowest
flow in the simulation) to cross the whole grid and for the accretion rate to show a repetitive pattern. The evolution
of $\dot{\bar M}_{\rm IBH}$ for the reference cases is shown in Fig.~\ref{accr}. As seen from the figures, the
interaction structure becomes more unstable; and the typical $\dot{\bar M}_{\rm IBH}$ decreases proportionally to the decreasing value of
$\theta$, that is, as the the medium-outflow interaction grows more direct. This result confirms what was already indicated in the simulations of \cite{li20}  and also suggested by \cite{bos20} regarding the
$\theta$-dependence of mechanical feedback. Averaging the QS-state $\dot{\bar M}_{\rm IBH}$ in
solid angle, we obtain an averaged ratio of $\dot{\bar M}_{\rm IBH}$ over the numerical accretion rate without outflow ($\dot{M}^{\rm num}_{\rm IBH}$) of:  
\begin{equation} 
<\xi^{\rm num}>=<\dot{\bar M}_{\rm IBH}>/\dot{M}^{\rm num}_{\rm IBH}\approx 0.4\,. 
\end{equation}


\begin{figure}        
       \centering       
       \includegraphics[width=6cm]{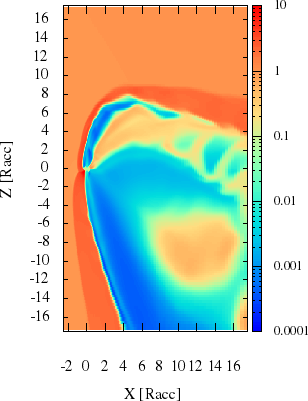}
       \includegraphics[width=8cm]{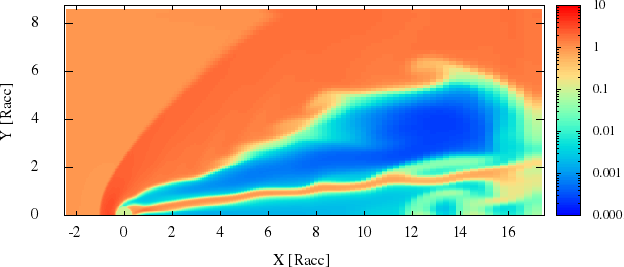}      
       \caption{Density map on the $y=0$ (top panel) and $z=0$ planes (bottom panel) for $\chi=0.4$, $\mathcal{R}=1/3$ and $\theta=45^\circ$, at $t_{\rm sim}\approx 75\,t_{\rm acc}$. The initial medium velocity is $10^7$~cm~s$^{-1}$, and the axis units $r_{\rm acc}\approx 8\times 10^{13}$~cm.}
       \label{densm7}
\end{figure}

\begin{figure}        
\includegraphics[width=8cm]{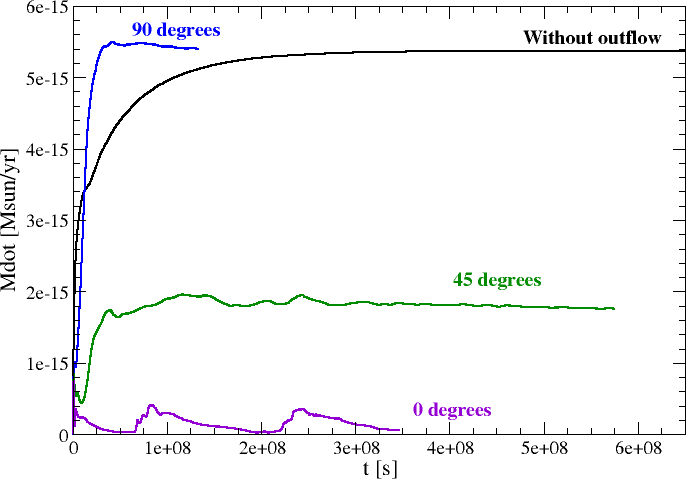}
\caption{Accretion rate for $v_{\rm IBH}=10^7$~cm~$^{-1}$ and $\theta=90^\circ$ (blue solid line), $\theta=45^\circ$ (green solid line) and $\theta=0^\circ$ (violet solid line). The case without outflow is also shown (black solid line). We remind that in these simulations $t_{\rm acc}\approx 8\times 10^6$~s.}
\label{accr2}
\end{figure}

Figure~\ref{densm7} presents the density maps in the $y=0$ and $z=0$ planes for $\theta=45^\circ$ and $v_{\rm IBH}=10^7$~cm~s$^{-1}$, and
the same cell size in units of $r_{\rm acc}$ as in the reference cases. For this intermediate $\theta$-value, the interaction
structure is mostly filled with shocked outflow material, whereas in the corresponding reference case, it was mostly filled by shocked medium material. The structure seems more unstable as well when it is far from the IBH sphere of influence. Density maps for the cases with
$\theta=90^\circ$ and $0^\circ$ are not shown as they resemble the corresponding reference cases. Figure~\ref{accr2} shows the
evolution of $\dot{\bar M}_{\rm IBH}$ for the three $\theta$-values, which is more affected by mechanical feedback than in the
reference cases, whereas the level of stability of the accretion curves is similar.
For this lower $v_{\rm IBH}$-value, the value of $<\xi^{\rm num}>$ approaches $\sim 0.2$ using this poorer sample of
$\theta$-values (although it is consistent with the value derived from the mentioned test trials with lower resolution). This result
indicates that the lower the value of $v_{\rm IBH}$, the stronger the effect of mechanical feedback on the IBH accretion rate.

\begin{figure}        
       \centering       
       \includegraphics[width=8cm]{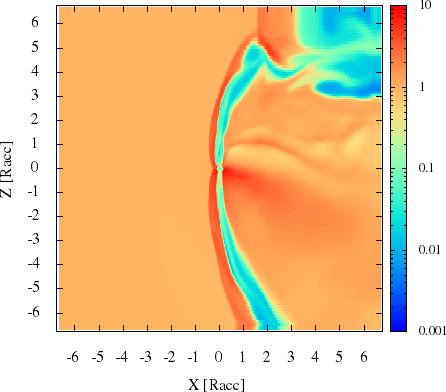}
       \includegraphics[width=8cm]{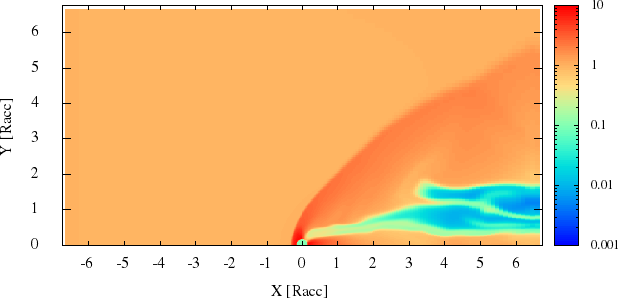}   
       \caption{Density map on the $y=0$ (top panel) and $z=0$ planes (bottom panel) for $\chi=0.4$, $\mathcal{R}=1/6$, and $\theta=60^\circ$, at $t_{\rm sim}\approx 12\,t_{\rm acc}$ in the high-resolution case. The initial medium velocity is $5\times 10^7$~cm~s$^{-1}$, and the axis units $r_{\rm acc}\approx 3\times 10^{12}$~cm.}
       \label{densm8}
\end{figure}

\begin{figure}        
       \centering       
       \includegraphics[width=8cm]{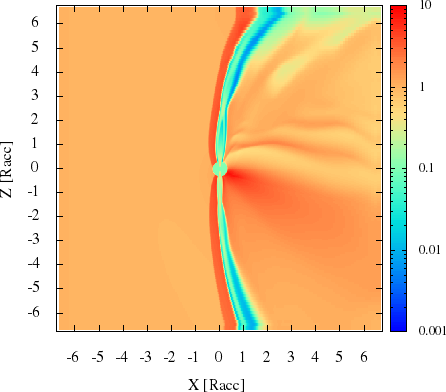}
       \includegraphics[width=8cm]{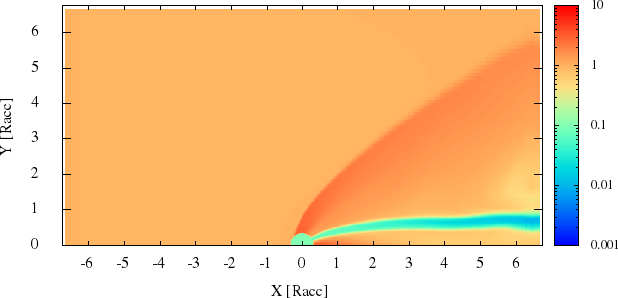}   
       \caption{Same as in Fig.~\ref{densm8} but for $\mathcal{R}=1/3$ and $\chi=0.2$.}
       \label{densm9}
\end{figure}

\begin{figure}        
       \centering       
       \includegraphics[width=8cm]{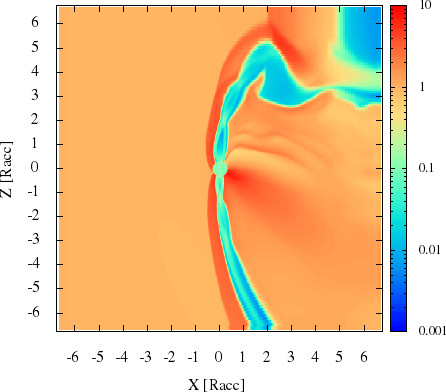}
       \includegraphics[width=8cm]{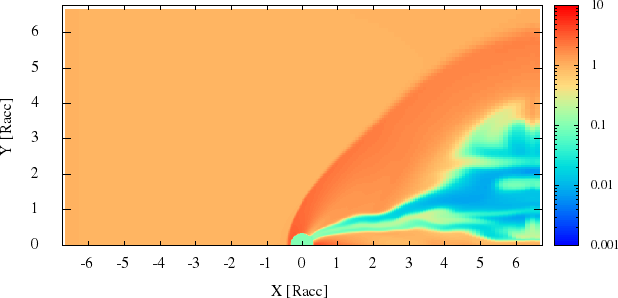}     
       \caption{Same as in Fig.~\ref{densm8} but for $\mathcal{R}=1/3$ and $\chi=0.4$.}
       \label{densm10}
\end{figure}

\begin{figure}        
\includegraphics[width=8cm]{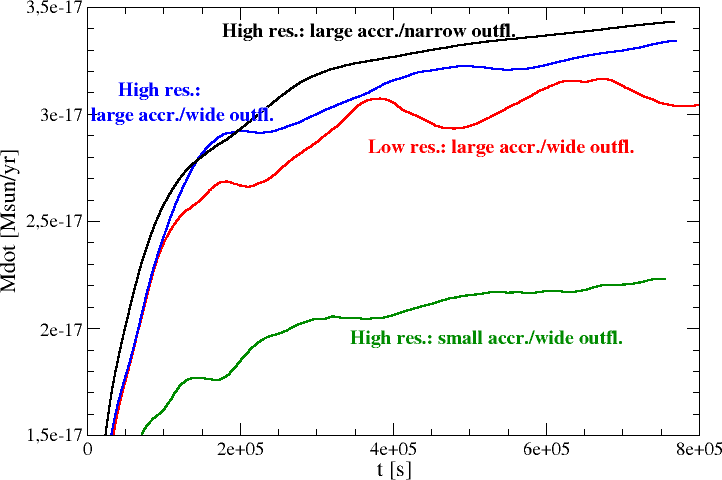}
\caption{Accretion rate for $v_{\rm IBH}=5\times 10^7$~cm~$^{-1}$ and $\theta=60^\circ$ in the high-resolution cases with $\mathcal{R}=1/3$ and $\chi=0.2$ (black solid line), $\mathcal{R}=1/3$ and $\chi=0.4$ (blue solid line), and $\mathcal{R}=1/6$ and $\chi=0.4$ (green solid line). The value of $t_{\rm sim}$ is $\approx 12\,t_{\rm acc}$. The reference case with the same $\theta$ is also shown (red solid line).}
\label{accr3}
\end{figure}

\begin{figure}        
       \centering       
       \includegraphics[width=8cm]{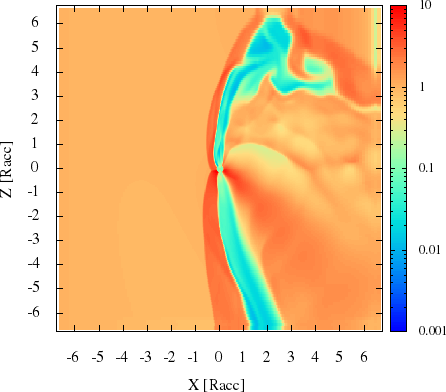}
       \includegraphics[width=8cm]{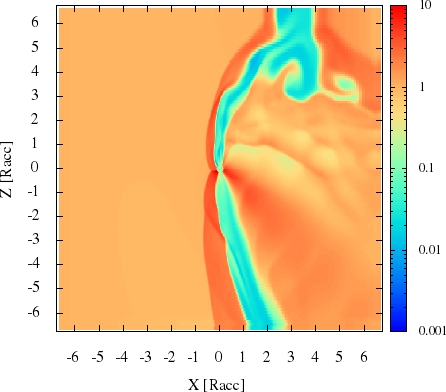}
       \includegraphics[width=8cm]{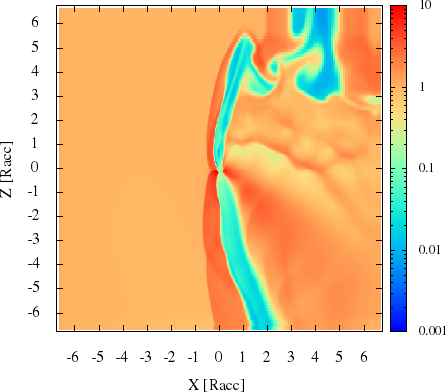}
       \caption{Same as in Fig.~\ref{densm8} but for $\theta=45^\circ$ and only the $y=0$ plane at three different times: $\approx 21$ (top panel), 22 (middle panel), and $23\,t_{\rm acc}$ (bottom panel).}
       \label{densm11}
\end{figure}

\begin{figure}        
\includegraphics[width=8cm]{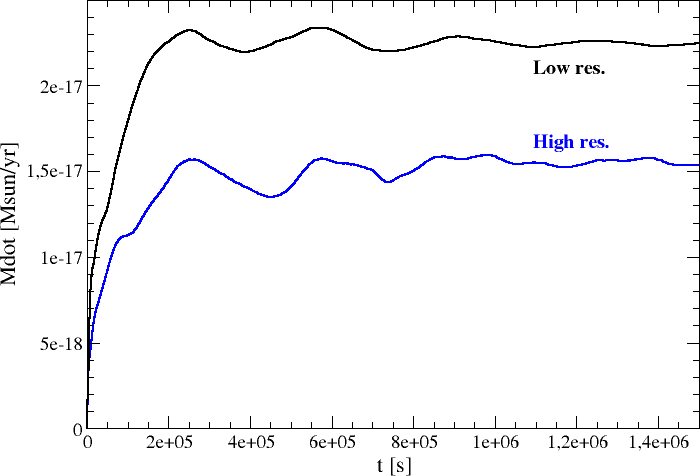}
\caption{Same as in Fig.~\ref{accr3} but for $\theta=45^\circ$, and $\chi=0.4$ and $\mathcal{R}=1/6$, with $t_{\rm sim}\approx 23\,t_{\rm acc}$ (blue solid line). The reference case (low-resolution) for the same $\theta$-value is also shown (black solid line).}
\label{accr4}
\end{figure}

Figures~\ref{densm8}--\ref{densm10} show density maps in the $y=0$ and $z=0$ planes for $\theta=60^\circ$, again with $v_{\rm
IBH}=5\times 10^7$~cm~s$^{-1}$, but now with twice the resolution of the reference cases. Two accretor sizes ($\mathcal{R}=1/6$ and
$1/3, $ as in Figs.~\ref{densm8} and \ref{densm10}) fixing $\chi=0.4$, and outflow half-opening angles ($\chi=0.2$ and 0.4, as in
Figs.~\ref{densm9} and \ref{densm10}) fixing $\mathcal{R}=1/6$, were explored. Figure~\ref{accr3} shows the evolution of $\dot{\bar
M}_{\rm IBH}$ for these three cases and the corresponding reference case. Figure~\ref{densm11} presents the same as Fig.~\ref{densm8}
($v_{\rm IBH}=5\times 10^7$~cm~s$^{-1}$, $\mathcal{R}=1/6$ and $\chi=0.4$), but now for $\theta=45^\circ$ and only the $y=0$ plane. To
illustrate the variability of the structure in the QS state, stronger than in the corresponding reference case, this figure
shows three different simulation times towards $t_{\rm sim}\approx 23\,t_{\rm acc}$, each separated by $\approx t_{\rm acc}$. The
evolution of $\dot{\bar M}_{\rm IBH}$ for this case and for the corresponding reference case is shown in Fig.~\ref{accr4}. The accretion
curves of $\theta=60^\circ$ and $45^\circ$ do not reflect a higher variability of $\dot{\bar M}_{\rm IBH}$ in the
high-resolution cases when compared to the reference ones.

The high-resolution simulations show that increasing the resolution by a factor of 2, while keeping the accretor size and the
outflow opening angle equal, yields similar results to the low-resolution simulations. On the other hand, making the resolution
twice higher, while keeping the accretor size and reducing the outflow half-opening angle by half,  somewhat reduces the impact of
mechanical feedback, hinting at a moderate impact of $\chi$ on this effect; we note that for $\chi\gtrsim 1$, the results are expected to be closer to the case with $\theta=0$. However, the most prominent result is that increasing the
resolution by a factor of 2, while keeping $\chi=0.4$  and decreasing the accretor size to a half, reduces the
predicted QS-state accretion rate by $\approx 33$\%.

\begin{figure}        
\includegraphics[width=7.5cm]{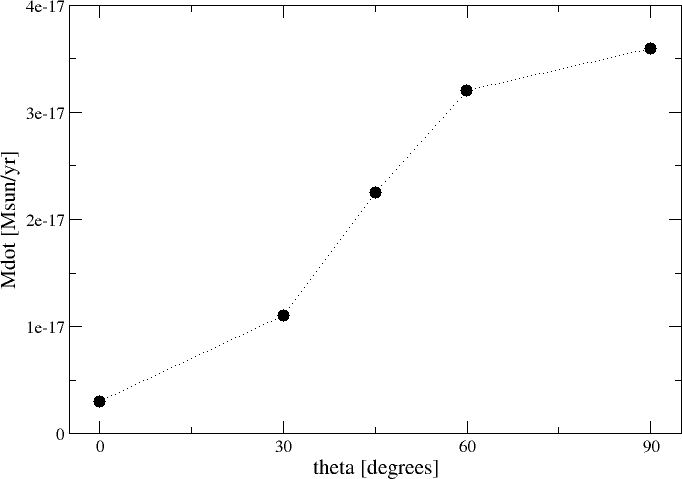}
\caption{Typical accretion rate versus $\theta$-value for the reference cases in the QS state.}
\label{anal1}
\end{figure}

\begin{figure}        
\includegraphics[width=7.5cm]{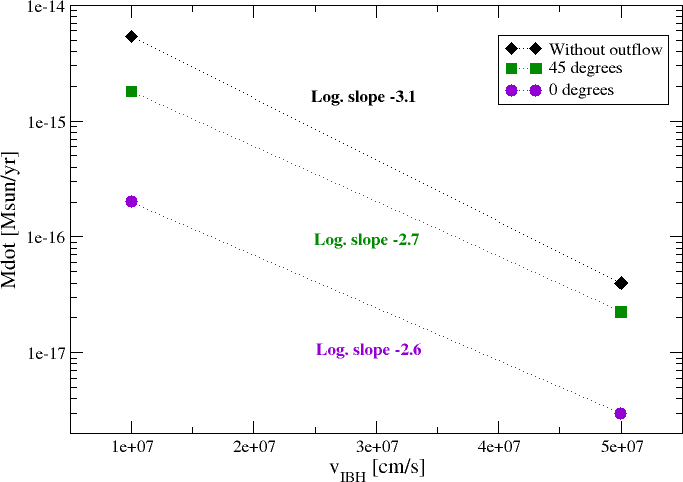}
\caption{Typical accretion rate versus $v_{\rm IBH}$ at $\theta=0^\circ$, $45^\circ$ and without outflow in the QS state. The slopes of a log-log representation of the lines are also given.}
\label{anal2}
\end{figure}

\begin{figure}        
\includegraphics[width=7.5cm]{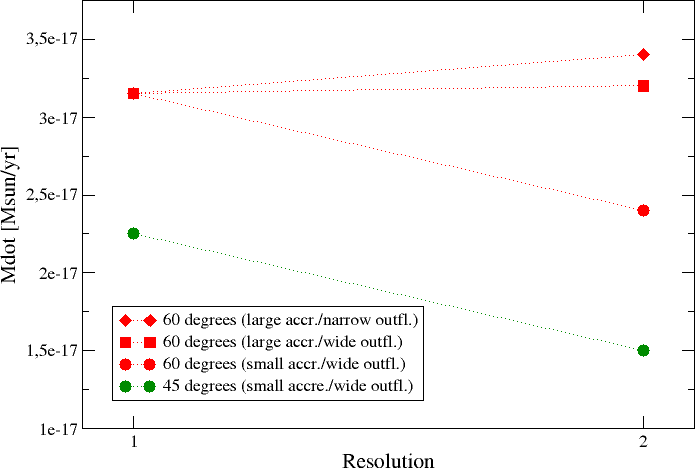}
\caption{Typical accretion rate versus degree of resolution (in the horizontal axis, 1: low; 2: high) for $v_{\rm IBH}=5\times 10^7$~cm~s$^{-1}$ and $\theta=45^\circ$ ($\mathcal{R}=1/6$, $\chi=0.4$) and $60^\circ$ ($\mathcal{R}=1/3$, $\chi=0.2$; $\mathcal{R}=1/3$, $\chi=0.4$; $\mathcal{R}=1/6$, $\chi=0.4$),  at $t_{\rm sim}\approx 12\,t_{\rm acc}$.}
\label{anal3}
\end{figure}

We conclude this section with some illustrative plots that summarize the obtained results. We represent the
typical accretion rate versus $\theta$-value for the reference cases in the QS state, and the result is shown in
Fig.~\ref{anal1}. The figure indicates that the stronger $\theta$-dependence of $\dot{\bar M}_{\rm IBH}$ is found at
intermediate $\theta$-values. In Fig.~\ref{anal2}, we present the typical accretion rate versus $v_{\rm IBH}$ at
$\theta=0^\circ$, $45^\circ$ and without outflow (very similar to the $90^\circ$ case) in the QS state. In that figure, the slopes of a log-log representation of
the lines are also given to better visualize the difference in behavior between the cases with outflow and the case
without. The slopes seem to indicate that the $v_{\rm IBH}$-dependence of $\dot{\bar M}_{\rm IBH}$ weakens somewhat when
the outflow is present. It is also worth noting that the case without outflow, with slope $-3.1$, is as expected pretty
close to the analytical value of $-3$ (see Eq.~\ref{eq:bondi_hoyle_accretion_rate}). Finally, in Fig.~\ref{anal3} we show
the typical accretion rate versus degree of resolution (low and high) for $v_{\rm IBH}=5\times
10^7$~cm~s$^{-1}$ and $\theta=45^\circ$ ($\mathcal{R}=1/6$, $\chi=0.4$) and $60^\circ$ ($\mathcal{R}=1/3$, $\chi=0.2$;
$\mathcal{R}=1/3$, $\chi=0.4$; $\mathcal{R}=1/6$, $\chi=0.4$),  at $t_{\rm sim}\approx 12\,t_{\rm acc}$. As noted in the
text, the stronger effect on $\dot{\bar M}_{\rm IBH}$ arises when the accretor radius is smaller, whereas the effect is similar for both $\theta$-angles.

\section{Summary and discussion}
\label{dis}

The results presented in this work demonstrate that the presence of outflows, either winds or jets, can easily interfere with
supersonic accretion through mechanical feedback on the scales of the IBH sphere of influence. As expected from the analysis
described in Sect.~\ref{phy}, the results confirm that both $\theta$ and $v_{\rm IBH}$ have a significant impact on the
accretion rate. The low-resolution simulations allow us to predict that a population of IBH with a random distribution of
outflow orientations and directions of motion will present a decrease in the averaged $\dot{\bar M}_{\rm IBH}$ over
$\dot{M}^{\rm num}_{\rm IBH}$ of $<\xi^{\rm num}>\sim 0.4$ and $0.2$ for $\eta_v$-values of $1/20$ and $1/100$, respectively.
However, the impact of resolution cannot be neglected. As shown by the high-resolution simulations with $\mathcal{R}=1/6$, which probed cases
with $\theta=60^\circ$ and $45^\circ$ and $v_{\rm IBH}=5\times 10^7$~cm~$s^{-1}$, $\xi^{\rm num}$ decreases by $\approx 33$\% with
respect to the corresponding reference cases for both $\theta$-values. Thus, if this resolution effect could be generalized to
other $\theta$- and $\eta_v$-values, our results would predict even lower values for $<\xi^{\rm num}>$, of $\sim 0.2-0.3$ and $\sim
0.1-0.2$ for $\eta_v$-values of $1/20$ and $1/100$, respectively. It is worth noting that if the medium were inhomogeneous on
spatial scales $\lesssim r_{\rm acc}$, fast variations in the accreted angular momentum and magnetic field geometry may lead to
IBH outflows changing direction on timescales $\lesssim t_{\rm acc}$. This could lead to some degree of isotropization of the
outflow effect on the medium, in which case $\xi$ would likely become smaller. Moreover, if instead of a sole IBH, the accretor would
consist of more than one object, several outflows with different orientations could form, which again may reduce $\xi$.

The mechanical feedback process described in this work, which has such a significant impact  on scales $\sim r_{\rm acc}$ is usually not considered when
estimating the mechanical and radiative outputs of IBH \citep[see however, e.g.,][where mechanical feedback is
discussed]{iok17,mat18,tak21,tak21b}. If outflows with enough momentum and power to reach beyond $r_{\rm acc}$ from the IBH are present
(which does not require much $L_{\rm o}$), mechanical feedback should affect the impact of supersonic IBHs on the environment, and
their detectability via electromagnetic radiation. An accurate evaluation of the impact of mechanical feedback, case by case, or
population-wise, would require us to account for many factors the assessment of which is outside the scope of this work, in particular: mass, spatial, and
velocity distribution of IBH; the medium density structure; the outflow production efficiency and velocity, and (though less significant)
half-opening angle. Nevertheless, despite a drop in $\dot{\bar M}_{\rm IBH}$ by a factor of several may seem rather minor
in front of so many uncertainties, mechanical feedback cannot be neglected, as such a drop in $\dot{\bar M}_{\rm IBH}$ appears to be a
sound prediction that goes against IBH detectability.

We conclude this work by noting that our results seem robust in their range of applicability, paving the way for simulations with higher resolution, with larger grids and longer durations, which should be carried out as a next step for even more general predictions.
The reasons behind this include: the fact that increasing the resolution further would allow for the exploration of smaller values of $\mathcal{R}$ and $\chi$; enlarging
the grid would allow the adoption of even lower $\eta_v$-values, or larger $\epsilon$-values; and more resourceful simulations would permit the study of
relativistic effects by adopting $v_{\rm o}$-values closer to $c$ without increasing $v_{\rm IBH}$ to compensate \citep[see][for a
dicussion on the impact of $\eta_v$ on the computing time]{bos20}. These simulations should be carried out in a supercomputing
facility with a code allowing for MPI-parallelization. In addition to all these improvements, we must nevertheless indicate that
the mode of accretion and outflow generation depends on the accretion rate in a complex way, with narrow jets and broad winds
alternatively dominating the outflows depending on the accretion regime \citep{bos20}. Therefore, as accretion-ejection phenomena
occur in the IBH vicinity, the study of mechanical feedback should also eventually account for the physics on spatial scales of both $\sim$ and $\ll r_{\rm acc}$. 


\section*{Acknowledgements}
We thank the anonymous referee for constructive and useful comments that helped to improve the manuscript.
V.B-R. acknowledges financial support from the State Agency for Research of the Spanish Ministry of Science and Innovation under grant PID2019-105510GB-C31 and through the ''Unit of Excellence Mar\'ia de Maeztu 2020-2023'' award to the Institute of Cosmos Sciences (CEX2019-000918-M). V.B-R. is Correspondent Researcher of CONICET, Argentina, at the IAR.

\bibliographystyle{aa}
\bibliography{biblio}


\end{document}